\documentclass[12pt,a4paper]{article}

\pdfoutput=1

\usepackage[pdftex]{graphics}
\usepackage{epsfig}
\usepackage{jheppub_modified}

\usepackage{mathrsfs}
\usepackage{amsmath, amsthm, amssymb}
\newcommand{\be}{\begin{equation}}
\newcommand{\ee}{\end{equation}}
\newcommand{\beq}{\begin{equation}}
\newcommand{\eeq}{\end{equation}}
\newcommand{\bea}{\begin{eqnarray}}
\newcommand{\eea}{\end{eqnarray}}
\def\simlt{\stackrel{<}{{}_\sim}}

\newcommand{\gsim}{\lower.7ex\hbox{$\;\stackrel{\textstyle>}{\sim}\;$}}
\newcommand{\lsim}{\lower.7ex\hbox{$\;\stackrel{\textstyle<}{\sim}\;$}}

\title{Fair scans of the seesaw. Consequences for predictions on LFV processes.}

\author[a]{J. Alberto Casas,}
\author[a]{Jes\'us M. Moreno,}
\author[b]{Nuria Rius,}
\author[b]{Roberto Ruiz de Austri}
\author[a]{and Bryan Zald\'ivar}
\affiliation[a]{ Instituto de F\'isica Te\'orica, IFT-UAM/CSIC \\
                 Nicolas Cabrera 15, UAM \\
                  Cantoblanco, 28049 Madrid, Spain }
\affiliation[b]{ Instituto de F\'isica Corpuscular, IFIC-UV/CSIC \\
        Apartado de Correos 22085, 46071 Valencia, Spain}
\emailAdd{alberto.casas@uam.es}
\emailAdd{nuria@ific.uv.es}
\emailAdd{bryan.zaldivar@uam.es}

\abstract{

\small
Usual analyses based on scans of the seesaw parameter-space can be biassed since they do not cover in a fair way the complete parameter-space. More precisely, we show that in the common ``$R$-parametrization", many acceptable $R$-matrices, compatible with the perturbativity of Yukawa couplings, are normally disregarded from the beginning, which produces biasses in the results. We give a straightforward procedure to scan the space of complex $R$-matrices in a complete way, giving a very simple rule to incorporate the perturbativity requirement as a condition for the entries of the $R$-matrix, something not considered before. As a relevant application of this, we show
that the extended believe that BR($\mu\rightarrow e,\gamma$) in supersymmetric seesaw models depends strongly on the value of $\theta_{13}$ is an ``optical effect" produced by such biassed scans, and does not hold after a careful analytical and numerical study. When the complete scan is done, BR($\mu\rightarrow e,\gamma$) gets very insensitive to $\theta_{13}$. Moreover, the values of the branching ratio are typically larger than those quoted in the literature, due to the large number of acceptable points in the parameter-space which were not considered before. Including (unflavoured) leptogenesis does not introduce any further dependence on $\theta_{13}$, although decreases the typical value of BR($\mu\rightarrow e,\gamma$).
}


\begin{document}
\maketitle

\section{Introduction}

There is a generalized believe  \cite{Raidal:2008jk},\cite{Antusch:2006vw} that, in the context of the supersymmetric (SUSY) seesaw scenario, the value of the neutrino mixing angle $\theta_{13}$ has a strong impact on Lepton Flavour Violation (LFV) processes, in particular on the branching ratio BR($\mu\rightarrow e,\gamma$). The basic idea is the following. 

As is well known, starting with universal conditions for the soft mass matrices (and thus implementing minimal flavour violation), the renormalization group (RG) running induces non-vanishing off-diagonal entries in the (left-handed) slepton mass matrix, $\mathbf{m_L}^2$. Such entries are mainly generated by the matrix of neutrino Yukawa couplings, ${\bf Y_\nu}$; more precisely $(\mathbf{m_L}^2)_{ij}\sim ({\bf Y^\dagger_\nu Y_\nu})_{ij}$. Then these off-diagonal entries enable LFV processes through one-loop diagrams \cite{Hisano:1995cp}. On the other hand, for given low-energy observables (i.e. neutrino masses and neutrino mixing matrix), the matrix ${\bf Y^\dagger_\nu Y_\nu}$ has plenty of freedom. This is because in the standard seesaw scenario there are more initial high-energy parameters (18) than low-energy observables (9). Consequently, for given low-energy observables the rate of LFV processes can vary within a certain range. The claim of ref. \cite{Antusch:2006vw} is that this range typically shifts to larger values as the $\theta_{13}$ mixing angle increases; furthermore, such behaviour is strengthened by leptogenesis constraints. The effect is apparently very important, producing increases of several orders of magnitude in BR($\mu\rightarrow e,\gamma$) as $\theta_{13}$ grows within its experimental window. This remarkable behaviour has been shown up mainly by performing scans upon the seesaw parameters (with and without leptogenesis constraints). An analytical explanation for it has been offered in \cite{Arganda:2006vw}.

In this paper we will show that these results are essentially an optical effect produced by scanning only a part of the whole parameter space of the seesaw. As we will see, when the whole parameter space is considered, the impact of $\theta_{13}$ is very small, even negligible. We will also show that, typically, the scans of the seesaw parameters in the literature are both incomplete and biassed because they do not consider the whole freedom in the parameter space compatible with the requirement of perturbativity. Such lack of completeness is more important for some parameters than for others, thus the bias. We will present easy rules to perform complete explorations incorporating the perturbativity requirement. The mentioned common bias is the main reason for the observed conspicuous dependence on $\theta_{13}$. 

In section 2 we present the framework and set the notation. Section 3 is devoted to describe the two parameterizations of the seesaw we are dealing with in this work. In section 4 we show and explain the apparent contradiction between the two approaches, concerning the dependence of BR($\mu\rightarrow e,\gamma$) on the $\theta_{13}$ angle.
In section 5 we give a procedure to scan the seesaw parameter space in a fair way, incorporating the constraint of perturbativity (details are given in Appendix A). In the same section and section 6, we study both in an analytical and numerical way  the dependence of BR($\mu\rightarrow e,\gamma$) on $\theta_{13}$, making use of the mentioned scan. Section 7 is devoted to the inclusion of leptogenesis constraints. Finally, in section 8 we present our main conclusions.

\section{Framework and notation}

From now on we use the conventions and notation of ref. \cite{Casas:2001sr}.

In the standard SUSY seesaw the relevant superpotential is
\bea
\label{Wseesaw}
{W}\ \  \supset\ \ e_R^{c\ T} {\bf Y_e} L\cdot  H_1\  +\ 
\nu_R^{c\ T} {\bf Y_\nu} L\cdot H_2\ -\  {1\over 2}\nu_R^{c\ T} {\bf M} \nu_R^c\  
,
\eea
where $L$ ($e_R^c$) are the leptonic doublets (charged singlets), $\nu_R$ are the right-handed neutrinos and $H_{1,2}$ are the two supersymmetric Higgs doublets. ${\bf Y_e}$ and ${\bf Y_\nu}$ are Yukawa matrices in flavor space (flavor indices are dropped) and ${\bf M}$ is the Majorana mass matrix of right-handed neutrinos. Once right-handed neutrinos are decoupled (at the seesaw scale $\sim { M}$) a mass operator is left in the effective superpotential,
\bea
\label{Weinberg}
{W}_{\rm eff}\ \  \supset\ \ e_R^{c\ T} {\bf Y_e} L\cdot \bar H_1\  +\ 
\frac{1}{2}({\bf Y_\nu}L H_2)^T {\bf M^{-1}} ({\bf Y_\nu}L H_2)\ .
\eea
Then, the effective mass matrix for the light ($\sim$ left-handed) neutrinos is
\bea
\label{Mnu}
{\cal M}_\nu = \langle H^0_2\rangle^2 {\bf \kappa}\ ,
\eea
with
\bea
\label{kappa}
{\bf \kappa} = {\bf Y_\nu}^T {\bf M}^{-1}{\bf Y_\nu}  ~.
\eea
Note that the seesaw equations (\ref{Weinberg}), (\ref{kappa}) are valid at the seesaw scale. Besides, they are obtained using the (reasonable) approximation of decoupling at a unique threshold, instead of a (more accurate) decoupling in three steps (the three right-handed neutrino masses).

The neutrino mass-eigenvalues, $m_i=\langle H^0_2\rangle^2 \kappa_i$, and the mixing matrix, $U_{\rm MNS}$, are given by
\bea
D_{\bf \kappa}\ = U_{\rm MNS}^T\ {\bf \kappa}\ U_{\rm MNS}, 
\;\;\;\;\;\;\;D_\kappa\equiv{\rm diag}(\kappa_1, 
\kappa_2, \kappa_3),
\eea
with $\kappa_1\leq \kappa_2\leq \kappa_3$. The standard parametrization of the MNS matrix is
\bea 
\hspace{-0.5cm}
U_{\rm MNS}=
\left(
\begin{array}{ccc}
c_{13}c_{12} & c_{13}s_{12} & 
s_{13}e^{-i\delta}\\
-c_{23}s_{12}-s_{23}s_{13}c_{12}e^{i\delta} & 
c_{23}c_{12}-s_{23}s_{13}s_{12}e^{i\delta} & s_{23}c_{13}\\
s_{23}s_{12}-c_{23}s_{13}c_{12}e^{i\delta} & 
-s_{23}c_{12}-c_{23}s_{13}s_{12}e^{i\delta} &
c_{23}c_{13}
\end{array}
\right)
\left(
\begin{array}{ccc}
e^{-i\phi/2} & &\\
 & e^{-i\phi'/2} &\\
&&1
\end{array}
\right)~.
\label{UMNS}
\eea

From now on, for the sake of notation clarity, we will drop labels in the neutrino mixing and Yukawa matrices: $U\equiv U_{\rm MNS}$, ${\bf Y}\equiv {\bf Y_\nu} $. 

\section{Parameterizations of the Seesaw}

The high-energy seesaw Lagrangian, given by the superpotential (\ref{Wseesaw}), is determined by the entries of the ${\bf Y}$ and ${\bf M}$ matrices, which contain 18 independent parameters. On the other hand, there are 9 low-energy neutrino observables: the three neutrino masses, $\propto \kappa_i$, and the three mixing angles and the three phases contained in $U$. Hence, for given values of the low-energy observables, the freedom in the seesaw Lagrangian expands a 9-dimensional parameter space. There are two main ways of describing such space (or, in other words, of parametrizing our ignorance). We will call them the $R-$parametrization and the $V_L-$parametrization.

\subsection{$R-$parametrization}

It was shown in \cite{Casas:2001sr} that, for a given set of low-energy observables, $\kappa_i$ and $U$, the neutrino Yukawa matrix (at the seesaw scale) has the form:
\bea
\label{param1}
{\bf Y}=D_{\sqrt{M}} R D_{\sqrt{\kappa}}U^\dagger ~,
\eea
where $D_{\sqrt{M}}={\rm diag}\{\sqrt{M_i}\}$,  $D_{\sqrt{\kappa}}={\rm diag}\{\sqrt{\kappa_i}\}$, and $R$ is a complex orthogonal ($3\times 3$) matrix. So, the 9 see-saw parameters that parametrize our ignorance are the three right-handed masses and the 3 complex angles defining $R$. A usual parametrization of $R$ is
\bea
\label{R}
R\ =\ 
\left(
\begin{array}{ccc}
c_{2}c_{3} ~& -c_{1}s_{3}-s_{1}s_{2}c_{3} ~& 
s_{1}s_{3}-c_{1}s_{2}c_{3}\\
c_2 s_3  ~& c_1 c_3 - s_1 s_2 s_3 ~& -s_1 c_3-c_1 s_2 s_3 \\
s_2 ~& s_1 c_2 ~& c_1 c_2
\end{array}
\right)~,
\eea
where $s_i$ ($c_i$) are the sine (cosine) of the three complex angles $\theta_i$. Eq.(\ref{R}) is general up to reflections changing the sign of $\det R$.

\subsection{$V_L-$parametrization}

An alternative to the $R-$parametrization is the so-called $V_L-$parametrization, see ref. \cite{Davidson:2002qv}. For a given set of low-energy observables, $\kappa_i$ and $U$, the neutrino Yukawa matrix (at the seesaw scale) can be written as
\bea
\label{param2}
{\bf Y}=V_R D_{Y} V_L^\dagger ~.
\eea
Here $D_{Y}$ is the diagonal matrix containing the three (real and positive) neutrino Yukawa couplings, $y_i$; $V_L$ is a unitary matrix with identical structure as the MNS matrix [eq.(\ref{UMNS})], but, of course, with three different mixing angles and three different phases; and $V_R$ has also identical structure but with the diagonal matrix of phases acting from the left (for more details see e.g. \cite{Casas:2006hf}). Here, the 9 independent parameters that parametrize our ignorance are the three Yukawa couplings, $y_i$, and the three angles and three phases contained in $V_L$. The $V_R-$matrix and the three right-handed neutrino masses are obtained by substituting eq.(\ref{param2}) in the seesaw expression (\ref{kappa}), namely
\bea
\label{VRM}
V_R^\dagger D_{M} V_R^*\ =\ D_Y V_L^\dagger U D_\kappa^{-1} U^T V_L^*  D_Y \ ,
\eea
i.e. $V_R$ is the unitary matrix that diagonalizes the symmetric matrix in the right hand side of (\ref{VRM}) and $D_M$ is the corresponding diagonal matrix, which contains the three right-handed masses.

\section{The impact of $\theta_{13}$}

As mentioned in the introduction, even starting with diagonal and universal soft masses, the RG running generates off-diagonal entries in them. In particular, the (left) slepton mass matrix gets off-diagonal entries $(\mathbf{m_L}^2)_{ij}$, $i\neq j$, proportional (in the leading-log approximation) to the $({\bf Y^\dagger Y})_{ij}$ matrix element. On the other hand, at first order in the mass-insertion expansion, the branching ratio of the LFV process $l_i\rightarrow l_j,\gamma$ (with $l_{i,j}$ charged leptons of the $i, j$ families) is non-vanishing and proportional to the squared of the corresponding off-diagonal entry, $|(\mathbf{m_L}^2)_{ij}|^2$.

Consequently, the dependence of, say BR($\mu\rightarrow e,\gamma$), on the MNS matrix (and in particular on $\theta_{13}$) occurs mainly via the dependence of $({\bf Y^\dagger Y})_{21}$ on it. 
This becomes more clear from the following approximate formula for the branching ratio, which will be useful later for qualitative discussions,
\begin{equation}
 \mathrm{BR}(\mu\rightarrow e,\gamma) \sim
\dfrac{\alpha^3}{G_F^2 m_S^8}\left|-\dfrac{1}{8\pi^2}(3m_0^2+A_0^2)\mathrm{log}\dfrac{M_X}{M} \right|^2
\left|({\bf Y^\dag Y})_{21} \right|^2 \tan^2\beta\ .
\label{BRaprox}
\end{equation}
Here $m_S$ represents a typical supersymmetric leptonic mass, and $m_0$, $A_0$ are the universal scalar mass and the universal trilinear coupling at the unification scale $M_X$.

 Next, we analyze the dependence of $({\bf Y^\dagger Y})_{21}$ on $\theta_{13}$ using the two parametrizations of the seesaw discussed in sect. 3. We will find first a kind of ``paradox", then we will discuss its explanation.

\subsection{A ``paradox"}

In the $R-$parametrization the ${\bf Y^\dagger Y}$ matrix can be easily obtained from eq.(\ref{param1}):
\bea
\label{YYR}
{\bf Y^\dagger Y}=U D_{\sqrt{\kappa}} R^\dagger D_{M} R D_{\sqrt{\kappa}}U^\dagger ~.
\eea
This equation tells us that, for given right-handed masses ($D_{M}$) and a given $R-$matrix, ${\bf Y^\dagger Y}$ (and thus LFV processes) has a non-trivial dependence on $U$. Let us concentrate for the moment on the $({\bf Y^\dagger Y})_{21}$ matrix element, which is the relevant one for $\mu\rightarrow e,\gamma$. Assuming a hierarchical spectrum of neutrinos, $\kappa_1\ll\kappa_2\ll \kappa_3$, we can expand $({\bf Y^\dagger Y})_{21}$ in powers of $\sqrt{\kappa_i}$. The first term of such expansion is
\bea
\label{YYR12}
({\bf Y^\dagger Y})_{21}\ =\ \kappa_3 U_{23} \left[R^\dagger D_{M} R\right]_{33} U^*_{13} \ +\ \cdots 
\eea
Since $|U_{13}| = |s_{13}|$ and $|U_{23}|\sim 1/\sqrt{2}$, we see that, in this approximation, 
\bea
\label{YYR12_2}
({\bf Y^\dagger Y})_{21}\ \propto\ s_{13} ~.
\eea
This dependence on $s_{13}$ is quite strong, and is really the source of the dependence of BR($\mu\rightarrow e,\gamma$) on $\sim s_{13}^2$ observed in the literature (this was also noticed in \cite{Arganda:2006vw}). An analogous argument shows that BR($\tau\rightarrow e,\gamma$) has a similar dependence on $s_{13}$, while BR($\tau\rightarrow \mu\gamma$) is almost independent of $\theta_{13}$. We will discuss soon the validity of the previous expansion, and thus of these results.

Let us now turn to the $V_L-$parametrization. From (\ref{param2})
\bea
\label{YYVL}
{\bf Y^\dagger Y}= V_L D_{Y}^2 V_L^\dagger ~.
\eea
Clearly, now ${\bf Y^\dagger Y}$ does not depend at all on $U$. For given Yukawa couplings ($D_Y$) any choice of $V_L$ is compatible with any choice of $U$ and thus of $\theta_{13}$. So, varying $\theta_{13}$ does not affect ${\bf Y^\dagger Y}$ at all. This result seems to be in contradiction with the one obtained using the $R-$parametrization. 

One might argue that changing $\theta_{13}$ in the $V_L-$parametrization means moving along a line of constant $y_i$ and $V_L$ in the seesaw parameter space; while changing $\theta_{13}$ in the $R-$parametrization means moving along a line of constant $M_i$ and $R$. It may happen that $({\bf Y^\dagger Y})_{21}$ keeps constant along the first line but it increases along the second one. This is true, but even assuming this possibility there remains a conflict, as we are about to see.

Let us work first in the $V_L-$parametrization. Assuming hierarchical neutrino Yukawa couplings, $y_1^2\ll y_2^2\ll y_3^2$, it is obvious that the choice of $V_L$ that maximizes $({\bf Y^\dagger Y})_{21}$ in eq.(\ref{YYVL}) is
$|(V_L)_{13}|=|(V_L)_{23}|=1/\sqrt{2}$. Then
\bea
\label{YYVLmax}
\left({\bf Y^\dagger Y}\right)_{21}^{\rm max}= \frac{1}{2} y_3^2 ~.
\eea
As mentioned above, this value is available for any choice of $U$. Let us keep fixed all mixings and phases in $U$, except $\theta_{13}$. For each value of $\theta_{13}$, the ${\bf Y^\dagger Y}$ matrix remains the same, but $D_M$ and $V_R$ change according to eq.(\ref{VRM}). The corresponding ${\bf Y}$ matrix is given by (\ref{param2}). Suppose we choose 
$\theta_{13}= 3^o$ and then calculate $D_M$ and $V_R$, and write ${\bf Y}$. This matrix can be easily expressed in the $R-$parametrization. Namely, in eq.(\ref{param1}) we can straightforwardly solve $R$ in terms of ${\bf Y}$, $U$, $D_M$ and $D_\kappa$, which are known. We can wonder now what happens if we keep these values of $D_M$ and $R$ fixed, and vary $\theta_{13}$. 
The ${\bf Y}$ matrix will change according to (\ref{param1}), but the Yukawa eigenvalues, $y_i^2$ will not change, as it is obvious from eq.(\ref{YYR}). From the point of view of the $V_L-$parametrization the change in ${\bf Y}$ is due to a change in $V_R$ {\em and} $V_L$. Therefore $\left({\bf Y^\dagger Y}\right)_{21}$ departs necessarily from its maximum value (\ref{YYVLmax}) and can only decrease in magnitude. However, from the point of view of the $R-$parametrization, the approximate expression eq.(\ref{YYR12}) tells us that an increase of $s_{13}$ should reflect in an increase in the magnitude of $\left({\bf Y^\dagger Y}\right)_{21}$! So, at least we have found a choice of $R$ for which the impact of $\theta_{13}$ on $\left({\bf Y^\dagger Y}\right)_{21}$ [and thus of BR($\mu\rightarrow e,\gamma$)] goes exactly opposite that claimed in the literature. 

\subsection{The reason behind}

The solution to the previous conflict can be found by doing the above steps explicit. We start with a choice for the Yukawa eigenvalues, $D_Y$, and a $V_L$ matrix that maximizes 
$\left({\bf Y^\dagger Y}\right)_{21}$ in eq.(\ref{YYVL}). The corresponding $D_M$, $V_R$ can be obtained from 
eq.(\ref{VRM}). One can construct now the ${\bf Y}$ matrix from (\ref{param2}). Then, using the $R-$parametrization (\ref{param1}), it is straightforward to derive the $R-$matrix (say $\hat R$) that corresponds to this optimal choice:
\bea
\label{Rmax}
\hat R\ =\ D_{\sqrt{M}}^{-1} V_R D_Y V_L^\dagger U D_{\sqrt{\kappa}}^{-1}\ ~,
\eea
(it is funny to check that $\hat R$ is orthogonal indeed). If we keep now $\hat R$ and $D_M$ constant but change the MNS matrix, $U\rightarrow U'$ (e.g. by varying $\theta_{13}$), it is straightforward from eq.(\ref{YYR}) that the new ${\bf Y'^\dagger Y'}$ matrix reads
\bea
\label{YYp}
{\bf Y'^\dagger Y'}= U' U^\dagger V_L D_{Y}^2 V_L^\dagger\ U U'^\dagger=U' U^\dagger 
{\bf Y^\dagger Y} U U'^\dagger ~.
\eea
Obviously, $\left|({\bf Y'^\dagger Y'})_{21}\right|\leq \left|({\bf Y^\dagger Y})_{21}\right|$ (recall that $V_L$ was ``designed" to maximize this quantity in eq.(\ref{YYVL})). So something goes wrong with the argument used to obtain eqs.~(\ref{YYR12}), (\ref{YYR12_2}). To see what, we construct the $\hat R^\dagger D_M \hat R$ matrix that appears in the expansion (\ref{YYR12}): 
\bea
\label{RMR}
\hat R^\dagger D_M \hat R\ =\ D_{\sqrt{\kappa}}^{-1}\ U^\dagger Y^\dagger Y U\ D_{\sqrt{\kappa}}^{-1}\  ~.
\eea
Hence
\bea
\label{RMRij}
\left(\hat R^\dagger D_M \hat R\right)_{ij}\ \propto \frac{1}{\sqrt{\kappa_i\kappa_j}}\ .
\eea
The consequence is that all terms neglected in (\ref{YYR12}) are in principle as large as the first term: the expansion is not sensible. In Fig.~1 we have plotted $(|{\bf Y^\dagger Y})_{21}|^2$ against $\theta_{13}$, keeping $\hat R$, $D_M$ constant, for this particular example. As expected, the maximum occurs at $\theta_{13}= 3^o$ and for larger $\theta_{13}$ the matrix element decreases.

\begin{figure}[ht!]
\centering
\hspace{-1cm}
\includegraphics[width=0.7\textwidth]{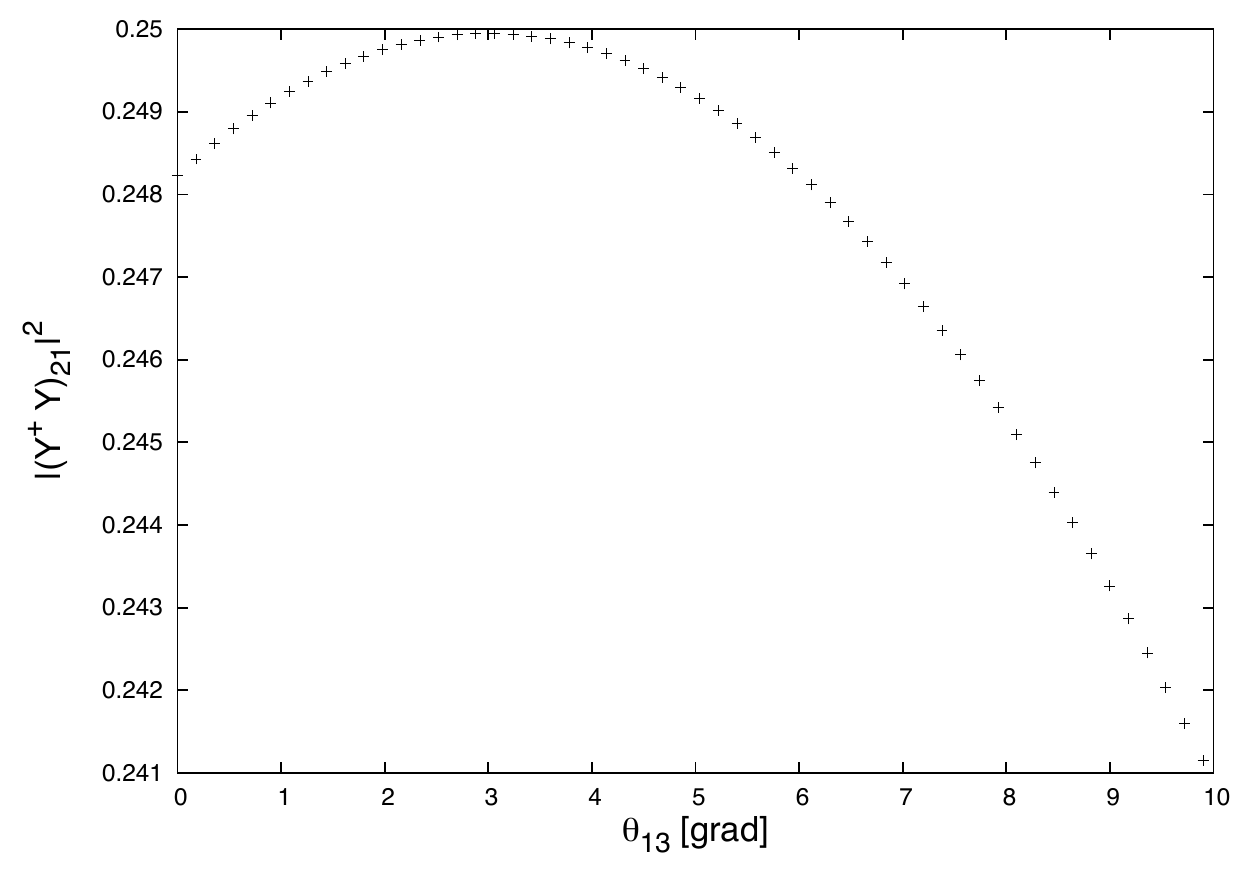}
\caption{ \small{ $(|{\bf Y^\dagger Y})_{21}|^2$ vs. $\theta_{13}$ for a given $\hat R-$matrix, correspondent the expression (4.7). Here $D_M=(10^{10},10^{11},10^{12})$ GeV.}  }
\label{fig1} 
\end{figure}

Thus we have constructed an explicit counter-example, where the dependence of $({\bf Y^\dagger Y})_{21}$ on $\theta_{13}$ goes opposite that naively expected. This raises the question: Is this ``wrong" behaviour a consequence of the special choice of the seesaw parameters ($\hat R$ and $D_M$) taken above or it is more general? What can we expect for a generic choice of $R$ and $D_M$?

\section{Fair scans in the $R-$matrix and the perturbativity condition}

The final questions of the previous section pose an interesting issue: how can we perform a truly generic scan in the $R-$matrix? In the literature it is common to separate the real and imaginary parts of the $\theta_i$ angles appearing in the parametrization (\ref{R}). Then the real part is varied as an ordinary angle, say $0\leq {\rm Re}\ \theta_i\leq 2\pi$, and the imaginary part is varied within a similar range. Note that such scan in ${\rm Re}\ \theta_i$ is really general, but the one in ${\rm Im}\ \theta_i$ is {\em not}. In principle ${\rm Im}\ \theta_i$ can take any value in the $\{-\infty, \infty\}$ range. Of course a too-large value of ${\rm Im}\ \theta_i$ is not realistic, since it leads to non-perturbative Yukawa couplings, making the whole approach inconsistent. In this section we examine what restrictions does the perturbativity criterion impose on the magnitude of the $R_{ij}$ entries. We will see that they are quite simple, but very different from just choosing a certain range for ${\rm Im}\ \theta_i$, which is the usual practice. 

The perturbativity requirement has to do with the Yukawa eigenvalues, $D_Y$. Since, for a given $D_M$ and $R$, these do not depend on the $U-$matrix [see eqs. (\ref{param1}) or (\ref{YYR})], the perturbativity criterion cannot depend on $U$ either. A simple and sensible approach is to impose a constraint on the trace of Yukawa couplings, say 
\bea
\label{Pert}
{\rm tr}\ {\bf Y^\dagger Y}\ =\ \sum_i y_i^2\  \simlt 3 ~,
\eea
(of course, any ${\cal O}(1)$ number is as good as 3 here). Now, from eq.(\ref{YYR}), 
\bea
\label{trYYR}
{\rm tr}\ {\bf Y^\dagger Y}= \sum_{j=1,2,3} \kappa_j \left[R^\dagger D_{M} R\right]_{jj}\ ,
\eea
so the perturbativity constraint (\ref{Pert}) translates into
\bea
\label{Pert_2}
|R_{ij}|^2 \ \simlt\ \frac{1}{M_i\kappa_j} ~.
\eea
This condition is very handy and easy-to-use. Besides, it clearly applies whether or not we consider a supersymmetric version of the seesaw. An important remark is that the perturbativity requirement does not affect equally the magnitude of the various $R_{ij}$ entries. Actually, they can be easily different by orders of magnitude. This is clearly in contrast with typical scans of the $R-$matrix in the literature. It also explains the structure of the $\hat R$-matrix [eq.(\ref{Rmax})]. Recall that $\hat R$ was constructed to maximize $\left({\bf Y^\dagger Y}\right)_{21}$ at a certain value of $\theta_{13}$. Then $\left({\bf Y^\dagger Y}\right)_{21}$ decreases for increasing $\theta_{13}$, in contrast to the usual behaviour observed in the literature. But $\hat R$ is not a usual matrix considered in the literature. Actually, it ``exploits" the perturbativity condition (\ref{Pert_2}) to the extreme, as is clear from eq.(\ref{Rmax}). But it still corresponds to a perfectly sensible ${\bf Y}$ matrix.

Now, we can pose the following question: For a given $D_M$, once $R$ is scanned in all its generality (respecting perturbativity and orthogonality conditions), what is the corresponding range for $\left({\bf Y^\dagger Y}\right)_{21}$ {\em and} how does it change when $\theta_{13}$ is varied? Intuitively, since $y_3^2\leq {\rm tr} {\bf Y^\dagger Y}$, we can expect a global range (see eq.(\ref{YYVLmax}))
\bea
\label{YYRmax}
0\lsim \left|({\bf Y^\dagger Y})_{21}\right|\lsim \frac{1}{2} {\rm tr} \bf Y^\dagger Y ~.
\eea
But still we do not know how feasible or natural is to reach these bounds depending on the value of $\theta_{13}$, or how $\left({\bf Y^\dagger Y}\right)_{21}$ changes with $\theta_{13}$ for a fixed vanilla $R$. We have studied this in a numerical way (more details in short), but we can get insight into these  issues by examining the expression of $\left({\bf Y^\dagger Y}\right)_{21}$ in the $R-$parametrization [eq.(\ref{YYR})] more closely. This expression can be written as
\bea
\label{YYB}
\left({\bf Y^\dagger Y}\right)_{21}\ =\ \sum_{j,k=1}^3 
U_{2k}\sqrt{\kappa_k}R_{jk}^* M_j 
\left(
R_{j3} \sqrt{ \kappa_3} U_{13}^*  \ +\ \sum_{i=1}^2 R_{ji} \sqrt{\kappa_i} U_{1i}^*  \right) ~.
\eea
Since $|U_{13}| = |s_{13}|$, the first term within the brackets is the responsible for the $\theta_{13}-$dependence observed in the literature. However, the other two terms within the brackets can be easily as big as the first one. Note first that $\sqrt{\kappa_2}$ is only a factor $\sim 1/\sqrt{6}$ smaller than $\sqrt{\kappa_3}$, a difference that is easily compensated by the fact that $|U_{13}|< |U_{12}|$. Besides, the perturbativity condition (\ref{Pert_2}) implies that typically all the $\sqrt{\kappa}$ factors in (\ref{YYB}) are compensated by the typical sizes of the $R-$matrix elements. In consequence, changing $\theta_{13}$ is not likely to have a noticeable impact on $\left({\bf Y^\dagger Y}\right)_{21}$, certainly not orders of magnitude for vanilla $R-$matrices.
This simple argument can be made more rigorous (and cumbersome) once the orthogonality conditions on the $R$ entries are imposed. But the basic result, that the range of $\left({\bf Y^\dagger Y}\right)_{21}$ cannot depend much on the value of $\theta_{13}$, remains. This contradicts the common lore in the literature, and it is much more consistent with the result obtained using the $V_L-$parametrization.

Let us now show the results of the numeric scan. First of all, we need a systematic procedure to scan the whole range of $R-$matrices, consistent with the perturbativity condition (\ref{Pert_2}) and the orthogonality condition, $R^T R={\bf 1}$. A simple way to do it is explained in Appendix A. Then, for each $R-$matrix considered, we scan $\theta_{13}$ in the $0^o-10^o$ range. Besides, for the numerical example we have taken the following values for the other parameters: 
\bea
\label{example}
m_1&=&10^{-12}~\mathrm{GeV}, \;\;\;\; m_2=9\times10^{-12}~\mathrm{GeV},\;\;\;\; m_3=5\times10^{-11}~\mathrm{GeV}, 
\nonumber\\
\theta_{12}&=&\pi/6, \;\;\;\; \theta_{23}= \pi/4,\;\;\;\; \delta=\phi_1=\phi_2=0, \;\;\; 
\nonumber\\
M_1&=&10^{10}~\mathrm{GeV}, \;\;\;\; M_2=10^{11}~\mathrm{GeV},\;\;\;\; M_3=10^{12}~\mathrm{GeV} ~.
\eea
Furthermore, we have used $\tan\beta\gsim 10$, so that $\langle H_2^0\rangle\simeq v/\sqrt{2}$, with $v=246$ GeV. The results of the scan in $R$ and $\theta_{13}$ are shown in Fig.~2. As expected, the dependence of $\left({\bf Y^\dagger Y}\right)_{21}$ on $\theta_{13}$ is very small, almost negligible. Note that, indeed, there are cases for which the dependence is stronger, corresponding to $R-$matrices which are far below the perturbativity limit (\ref{Pert_2}), but those are statistically rare exceptions.
\begin{figure}[ht!]
\centering
\hspace{-1cm}
\includegraphics[width=0.7\textwidth]{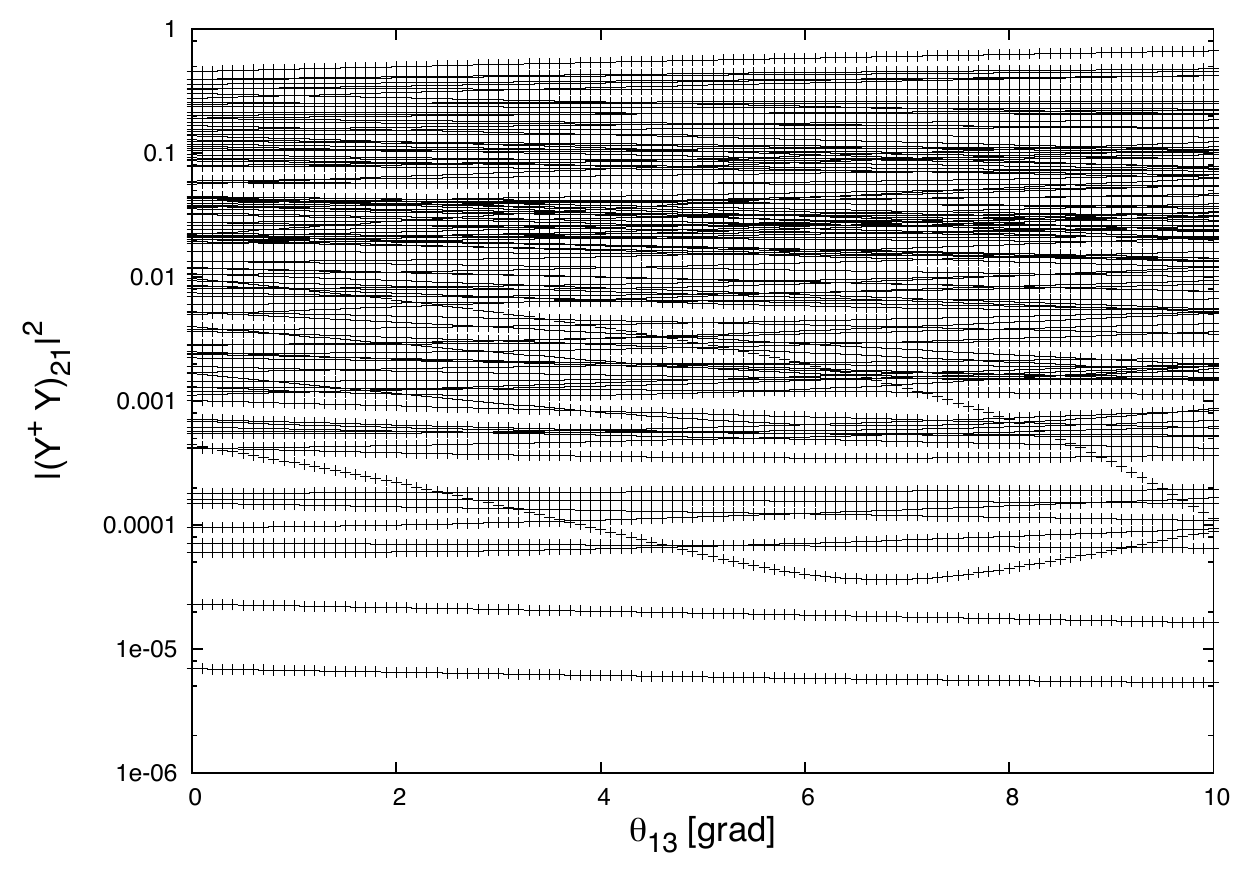}
\caption{ \small{ Scatter plot of  $(|{\bf Y^\dagger Y})_{21}|^2$ vs. $\theta_{13}$ for different $R-$matrices obeying the perturbativity condition (5.3). } }
\label{fig2} 
\end{figure}

\section{The branching ratio BR($\mu\rightarrow e,\gamma$)}

In oder to translate these results about $\left({\bf Y^\dagger Y}\right)_{21}$ into predictions for the branching ratio BR($\mu\rightarrow e,\gamma$), we have to assume first a supersymmetric model. We have chosen a minimal supergravity (mSUGRA) model defined by 
\bea
\label{mSUGRA}
m_0= 500 \, {\rm GeV}, \; M_{1/2}=250 \,{\rm GeV}, \; A_{0}=-100 \,{\rm GeV}, \;  \tan\beta= 10,
\eea
where $m_0$, $M_{1/2}$ and $A_0$ are the universal scalar mass, gaugino mass and trilinear coupling at the unification scale $M_X$. Then we have calculated BR($\mu\rightarrow e,\gamma$) for the same random set of $R-$matrices used for Fig. 2. The computation was made by means of an own modified version of the SPheno code \cite{Porod:2003um}, which uses the full one-loop expressions of ref. \cite{Pierce:1996zz}. The results are shown in Fig. 3 (left panel). As expected, the dependence of BR($\mu\rightarrow e,\gamma$) on $\theta_{13}$ follows closely the one of $\left|({\bf Y^\dagger Y})_{21}\right|^2$, shown in Fig. 2. Concerning the size of 
BR($\mu\rightarrow e,\gamma$), we see that in general is very large, quite above the experimental upper bound. This is in fact not surprising: using the approximate general range (\ref{YYRmax}) and the approximate formula (\ref{BRaprox}), one can check that the expected branching ratio is very large. Some comments are in order here. First, although it is common lore that BR($\mu\rightarrow e,\gamma$) can be quite large for a typical minimal SUGRA model, normally the actual values quoted in the literature are below those obtained here. This is because the scans in the $R-$matrix were not complete and many $R-$matrices compatible with perturbativity were not considered. Second, although it is not visible in Fig.3, there are of course choices of $R$ leading to branching ratios well below the experimental limit. What happens is that, scanning the space of the $R-$matrices in the way we did it, the number of those ``good" $R-$matrices is relatively very small.
We find this a very suggesting result. However, one has to keep in mind that scanning the $R-$parameter space in a different way (in Bayesian language, using a different prior for that space), the abundance of those ``good" $R-$matrices will change. The dependence of these results on the prior is out of the scope of this paper. Nevertheless it is worth noticing that the same prior dependence was implicit in the scatter plots shown in the previous literature. Let us finally remark that by changing the parameters (\ref{mSUGRA}) of the mSUGRA model, the branching ratio changes parametrically, as indicated in the approximate formula (\ref{BRaprox}).

For the sake of comparison between parametrizations, we have also shown in Fig. 3 (right panel) a similar survey using the $V_L-$parametrization. In this case we have to choose the values of the three neutrino Yukawa couplings. We have taken $y_1=0.0011$, $y_2=0.03$, $y_3= 1$. Besides, we have chosen the same mSUGRA model defined in eq.(\ref{mSUGRA}). Recall that in this parametrization $\left({\bf Y^\dagger Y}\right)_{21}$ does not depend at all on $\theta_{13}$. The branching ratio does, but in a marginal way. All this is apparent from Fig. 3. Now, comparing the surveys with the $R-$ and $V_L-$parametrizations (left and right panels) we note a similar insensitivity to $\theta_{13}$, which is one of our main results. Besides, in the figure we see that the branching ratios in the $R-$parametrization can be more than one order of magnitude larger than in the $V_L$ one. This is mainly due to the choice $y_3= 1$ in the latter. Note that in the $R-$parametrization we have imposed ${\rm tr}{\bf Y^\dagger Y}\leq 3$, which allows $y_3^2\lsim 3$. Therefore the upper limit of $\left|({\bf Y^\dagger Y})_{21}\right|^2$ can be almost one order of magnitude bigger than in this $V_L$ survey, see equations (\ref{YYRmax}), (\ref{YYRmax}). Furthermore, in the $V_L-$parametrization the righthanded neutrino masses are an output. In the example chosen they come out typically bigger (and less degenerate) than the choice made for the $R-$parametrization, eq.(\ref{example}). This pushes downwards further the branching ratio through the log factor, see eq.(\ref{BRaprox}). Hence, the two surveys are perfectly consistent.

\begin{figure}[ht!]
  \begin{center}
    \begin{tabular}{cc}
      \resizebox{70mm}{!}{\includegraphics{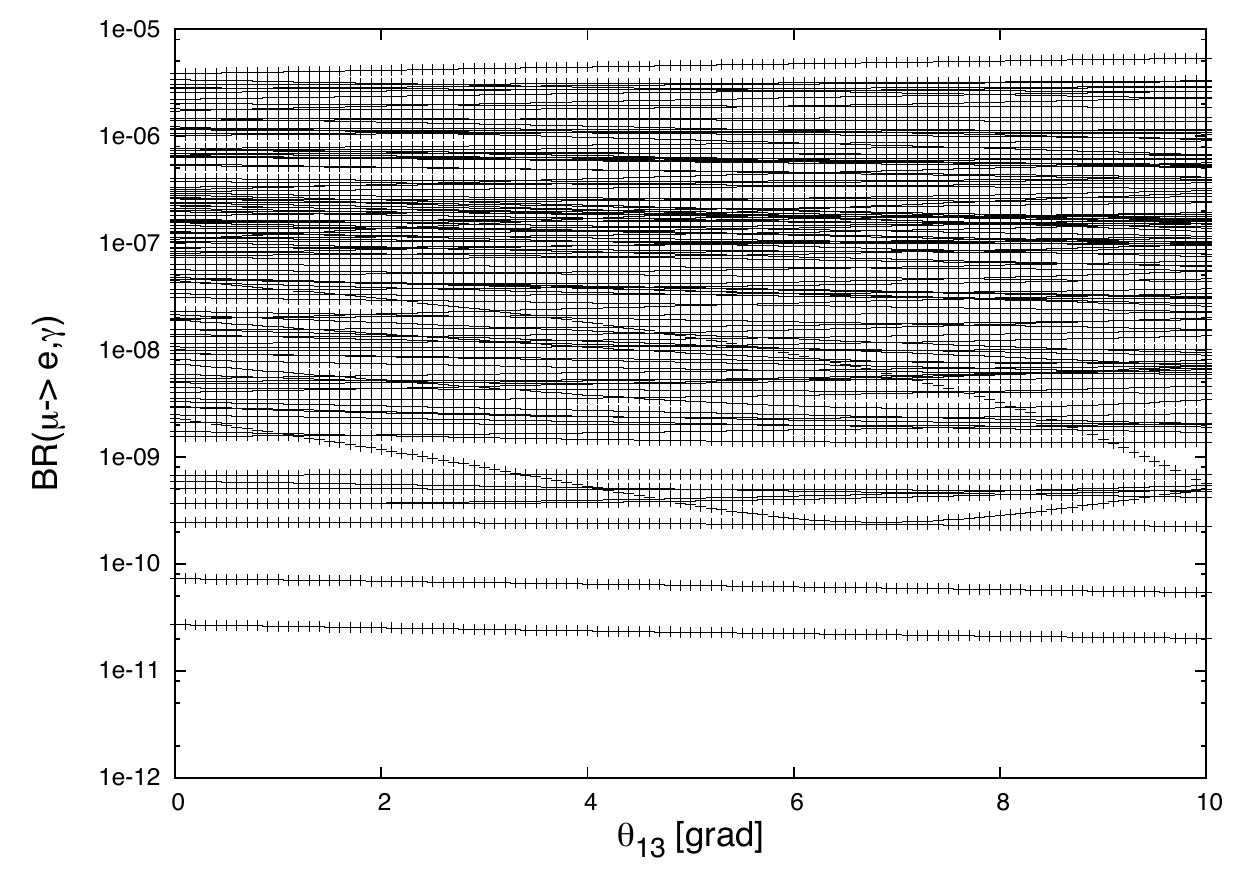}} &
      \resizebox{70mm}{!}{\includegraphics{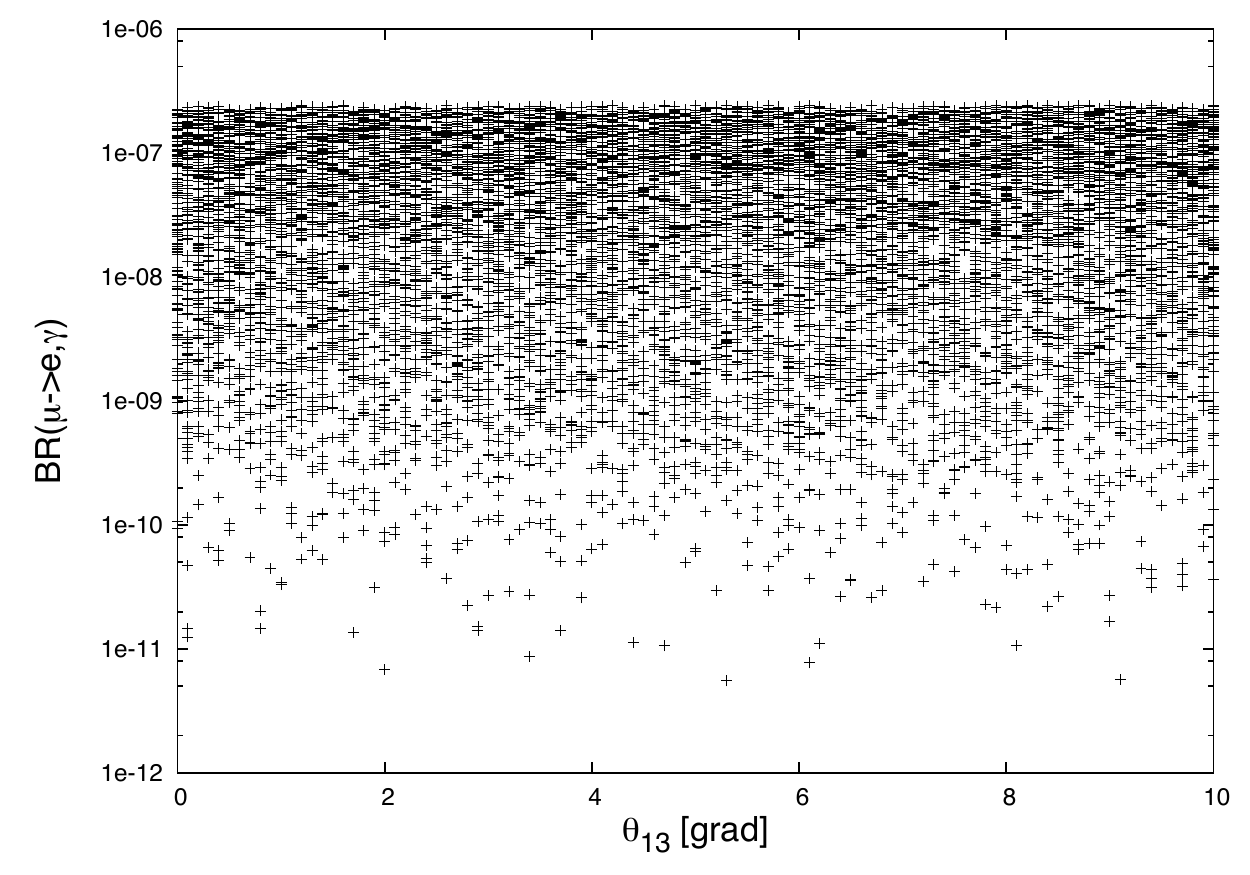}} \\
    \end{tabular}
    \caption{BR($\mu\rightarrow e,\gamma$) vs. $\theta_{13}$. (left) $R-$parametrization; (right) $V_L-$parametrization.}
    \label{BR-noleptog}
  \end{center}
\end{figure}

Of course, the results of this section could be different if the set of $R-$matrices (or $V_L$ and $V_R$ matrices for the $V_L$-parametrization) is constrained by additional considerations. This is the case of GUT models (as those studied in \cite{Masiero:2004js}), where non-trivial correlations can indeed occur. 

The last point raises a final question: what happens if leptogenesis constraints are imposed in the scenario? Actually, the impact of $\theta_{13}$ on LFV processes was reported to get reinforced  once successful leptogenesis is incorporated in the analysis. In the next section, we re-visit the leptogenesis issue.

\section{Constraints from leptogenesis}
\label{lepto}

The baryon asymmetry of the universe (BAU) is usually defined as the ratio of the number density of baryons $n_B$ to the number density of photons $n_\gamma$ \cite{Giudice:2003jh}. Its present experimental value is 
\cite{Komatsu:2010fb}
\begin{equation}
\dfrac{n_B}{n_\gamma} = (6.19\pm 0.15)\times10^{-10}.
\label{mainlepto}
\end{equation}
Perhaps the most popular mechanism to generate such BAU is nowadays thermal leptogenesis, in which a net lepton number is produced by the out-of-equilibrium decay of the (seesaw) right-handed neutrinos. Then the lepton number is converted into baryon number by sphaleron-mediated processes. Note here that the relevant Lagrangian, defined by the superpotential (\ref{Wseesaw}), contains the required lepton number and CP violating-terms (the latter are provided by appropriate phases in the ${\bf Y}$ matrix).

The final value for the BAU is given by:
\begin{equation}
\dfrac{n_B}{n_\gamma} = \dfrac{n_1}{n_\gamma}~ C_{\mathrm{sphal}} ~\epsilon~\eta ~,
\label{leptog1}
\end{equation}
where $n_1$ is the equilibrium number density of the lightest righthanded neutrino, which is the main responsible for the asymmetry for hierarchical righthanded masses, $M_1 \ll M_2\ll M_3$; $C_{\mathrm{sphal}}$ contains the sphaleron effect, $\epsilon$ is the CP-violating contribution to the asymmetry, and 
$\eta$ is the efficiency factor,  which takes into account the partial erasure of the CP asymmetry by inverse decays and scattering processes. In the minimal supersymmetric standard model eq.(\ref{leptog1}) can be written as \cite{Giudice:2003jh}
\begin{equation}
 \dfrac{n_B}{n_\gamma} \simeq -1.04\times10^{-2}~\epsilon~\eta ~.
\label{leptogmssm}
\end{equation}
The thermal production of righthanded neutrinos is suppressed unless $M_1\lsim T_R$, where $T_R$ is the reheating temperature after inflation. On the other hand, $T_R$ cannot be much larger than $10^{10}$ GeV, to avoid the gravitino problem. In the following we will assume $M_1\simeq T_R\simeq 10^{10}$ GeV. In this temperature regime, one or more charged-lepton mass-eigenstates $\ell$ ($\ell=e,\mu,\tau$) are in equilibrium in the thermal bath, and flavour effects can be significant because  the corresponding lepton asymmetries follow an independent evolution \cite{Abada:2006fw, Nardi:2006fx,Abada:2006ea}. 
However  we will not consider flavour effects here,  since our main goal in this paper is to examine the dependence of the results on $\theta_{13}$, 
and to compare the results with the previous literature (where flavor effects were not considered)\footnote{
It is worth mentioning here that in \cite{Davidson:2008pf} it was shown that successful leptogenesis is possible within the SUSY 
seesaw  for any value of the still unmeasured low energy neutrino parameters (including $\theta_{13}$), taking into account 
flavour effects and using the $V_L$ - parametrization. This may be an indication that flavour effects are not going to introduce any dramatic dependence on $\theta_{13}$.
}.

In the unflavored case, the efficiency factor, $\eta$, is given by
\begin{equation}
 \eta=\left[\left(\dfrac{2 (\tilde m_e +\tilde m_\mu + \tilde m_\tau)}{m_*}\right)^{1.16}+\left(\dfrac{2 m_*}{\tilde m_e +\tilde m_\mu + \tilde m_\tau}\right)\right]^{-1} ~,
\label{eta}
\end{equation}
where 
\[\tilde m_i = |{\bf Y}_{1i}|^2 v_u^2/M_1 \hspace{1cm} i=e,\mu,\tau
\]
\[m_* = 4\pi v_u^2 H_1/M_1^2 \sim 10^{-3} eV
\]
\begin{equation}
 H_1 =H(T=M_1) = 1.66 \sqrt{g_*}M_1^2/M_{\mathrm{Planck}} ~,
\label{whereUnflavored}
\end{equation}
\begin{equation}
 g_*|_{\mathrm{MSSM}} = 228.75\ ~,
\end{equation}
with $H$ denoting the Hubble parameter. The CP factor, $\epsilon$, is given by \cite{Covi:1996wh}:
\begin{equation}
 \epsilon= \dfrac{1}{8\pi}
\dfrac{\sum_{j=\mu,\tau} \mathrm{Im}\{[({\bf Y Y^\dag})_{1j}]^2\}~ g(M^2_j/M^2_1)}
{({\bf YY^\dag})_{11}} ~,
\label{epsilon}
\end{equation}
where $g(x)=\sqrt{x}\left[\dfrac{2}{1-x}-\mathrm{ln}\dfrac{x+1}{x}\right]$. 

Note that both $\eta$ and $\epsilon$ depend on the Yukawa couplings through the combination ${\bf YY^\dag}$, which, in the $R-$parametrization -eq.(\ref{param1})- is independent of $U$
\begin{equation}
{\bf Y Y^\dagger}= D_{\sqrt{M}} R D_{\kappa} R^\dagger   D_{\sqrt{M}}\ .
\label{YYdagR}
\end{equation}
Hence, the BAU, given by (\ref{leptogmssm}), does not depend on $\theta_{13}$. 

Still, it could happen that leptogenesis constraints strengthened the dependence of BR($\mu\rightarrow e,\gamma$) on $\theta_{13}$. As we have seen in the previous section, there are some $R-$matrices for which the dependence of $\left({\bf Y^\dagger Y}\right)_{21}$ (and thus of the branching ratio) on $\theta_{13}$ is important. If (in an extreme case)  those $R-$matrices were precisely those selected by successful leptogenesis, we would find a strong dependence of the branching ratio on $\theta_{13}$ in the complete scenario.
To analyze whether this possibility (or a more moderate one) takes really place, it is important to perform a complete scan of the $R-$matrix space. Performing partial scans in this space can be useful to show particular features, but it may introduce unwanted biasses: one could artificially select $R-$matrices that lead to strong dependences of BR($\mu\rightarrow e,\gamma$) on $\theta_{13}$ (as has happened in previous literature). 

\begin{figure}[ht!]
\centering
\hspace{-1cm}
\includegraphics[width=0.7\textwidth]{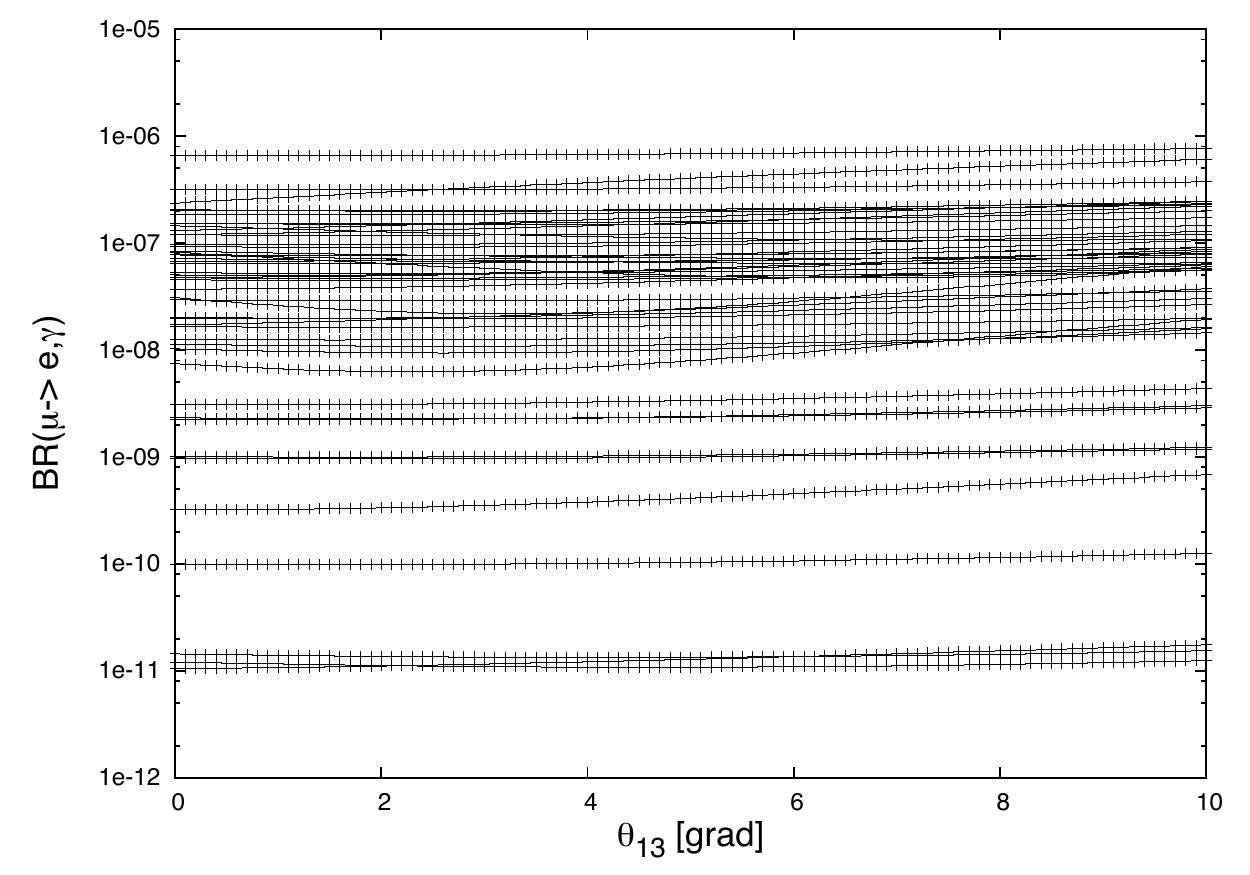}
  \caption{BR($\mu\rightarrow e,\gamma$) vs. $\theta_{13}$ in $R-$parametrization; including leptogenesis constraint.}
\label{fig4} 
\end{figure}

Fig. 4 shows the dependence of the BR($\mu\rightarrow e,\gamma$) on $\theta_{13}$ after imposing successful leptogenesis and scanning $R$ in its whole parameter space. Clearly, no important dependence on $\theta_{13}$ is observed. The trend is very similar to the one with no leptogenesis constraints (Fig. 3). The conclusion is that leptogenesis constraints do not enhance (or create) any dependence on $\theta_{13}$. 

Note that when the leptogenesis constraint is imposed, the values BR($\mu\rightarrow e,\gamma$) decrease at least one order of magnitude with respect to the case without leptogenesis, Fig.~3 (left). This behaviour can be understood
from eqs.~(\ref{leptogmssm}), (\ref{epsilon}) and (\ref{eta}). The baryon asymmetry depends on the neutrino Yukawa matrix {\bf Y} both through the CP asymmetry $\epsilon$  
and the efficiency factor $\eta$. Note from eq.~(\ref{epsilon}) that the CP asymmetry is mainly driven by the ${\nu_R}_2$ and ${\nu_R}_3$ Yukawa 
couplings (that is, the second and third rows of $R$), since the dependence on the ${\nu_R}_1$ Yukawas approximately cancels out due to the  $({\bf Y Y^\dag})_{11}$ factor in the denominator.
As a consequence, we find a mild dependence of $\epsilon$ on the elements in the first row 
of $R$ (see Fig.~4).

The efficiency factor, $\eta$, in eq.~(\ref{eta}) smoothly interpolates between the weak 
($\sum_{i}\tilde m_i  \ll m_* $) and strong ($\sum_{i}\tilde m_i  \gg m_* $) washout regimes. 
However for all the points in our scans we find that $\sum_{i}\tilde m_i \gg m_* $, thus 
the efficiency factor is given by
\begin{equation}
 \eta \approx \left[\frac{2 \sum_{i}\tilde m_i}{m_*}\right]^{-1.16},
\hspace{1cm} i=e,\mu,\tau \ ,
\label{eta2}
\end{equation}
and it becomes clear that $\eta$ decreases for larger ${\bf Y}_{1i}$, and thus for larger $R-$matrix elements, see eq.~(\ref{param1}). So, after imposing enough BAU, only those $R-$matrices which are sufficiently small in order to optimize $\eta$ are selected, thus reducing the final value of the branching ratio. 

\begin{figure}[ht!]
\centering
\hspace{-1cm}
\includegraphics[width=0.7\textwidth]{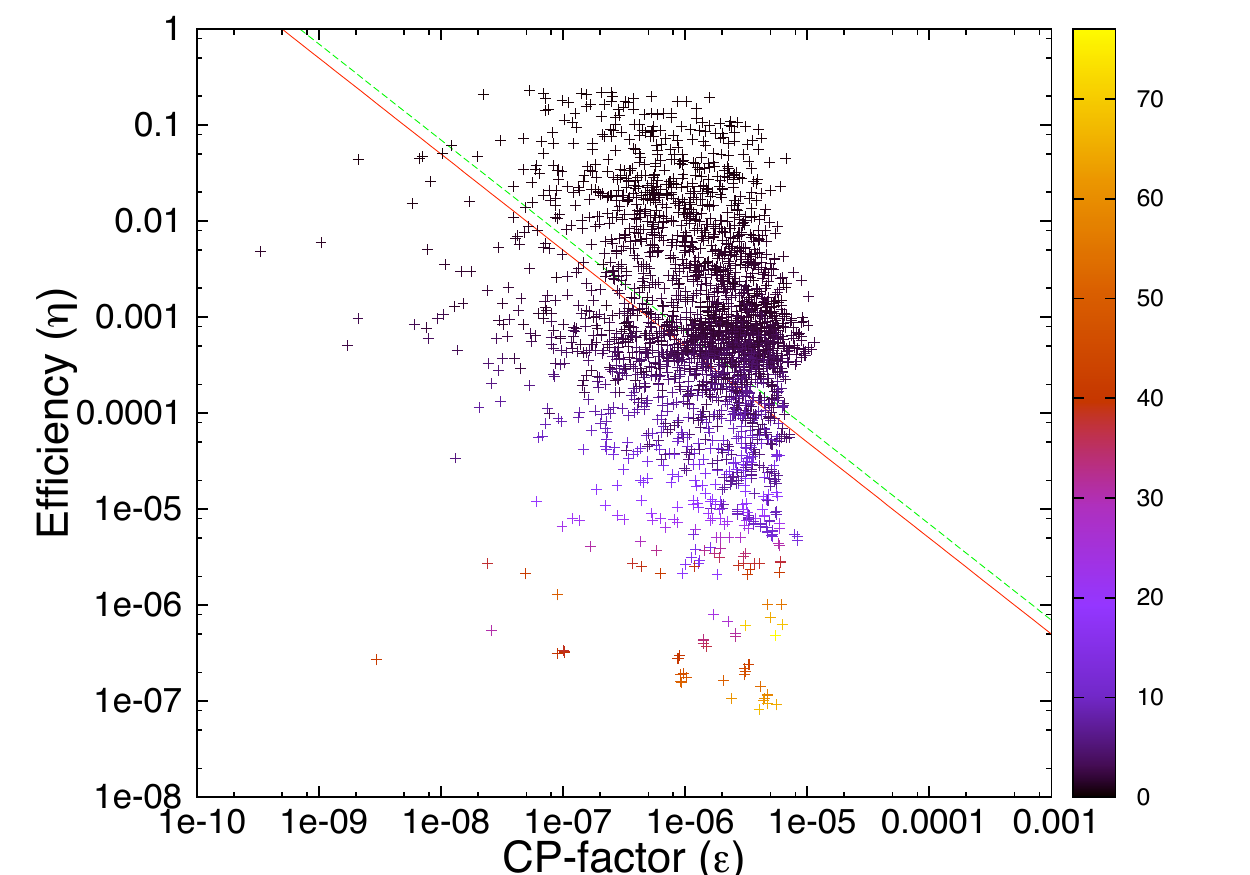}
\caption{ \small{Scatter plot of $\epsilon$ and $\eta$ for random R-matrices, every point represents a different $R-$matrix. The color coordinate represents the average absolute-value of the elements in the first row of $R$. The red and green lines denote the lower and upper BAU limits, respectively. For clarity in the plot, we have relaxed the allowed BAU window to be $[5-7]\times 10^{-10}$. } }
\label{fig5} 
\end{figure}

In Fig.~5 we  have performed a scatter plot showing the values of $\eta$ and $\epsilon$. The experimental window is also shown for reference. From the figure it is clear that enough BAU is only produced if the elements of the first row in $R$ are smaller than their perturbative bound, in agreement with our previous analytical estimations. This is the reason for the decrease of the branching ratio. Besides, there is no color gradient in the 
$\epsilon-$direction, which means that the same values of $\epsilon$ can be reached for any 
value of the first row entries of the  $R-$matrix. This is in contrast with the case of $\eta$, which changes dramatically when varying the typical size of the first row $R-$matrix elements. 

\section{Conclusions}

The main results and conclusions of this paper are the following:

\begin{itemize}

\item
The extended believe that the branching ratio BR($\mu\rightarrow e,\gamma$) in supersymmetric seesaw models depends strongly on the value of $\theta_{13}$ does not hold after a careful analytical and numerical study.

\item We have analyzed this issue using two alternative parametrizations of the 9 degrees of freedom that, besides the 9 low-energy observables (neutrino masses, mixings and phases), expand the parameter space of the seesaw scenario. This amounts to two alternative ways of traveling across this 9-dimensional space or, in other words, of parametrizing our ignorance. These are called the $R$-parametrization (sect. 3.1) and the $V_L$-parametrization (sect. 3.2)

\item The main potential dependence of BR($\mu\rightarrow e,\gamma$) on $\theta_{13}$ occurs through the ${\bf Y^\dagger_\nu Y_\nu}$ matrix, where ${\bf Y_\nu}$ is the neutrino Yukawa matrix. In the $V_L$-parametrization, this quantity is trivially insensitive to $\theta_{13}$ (or to any observable parameter), so BR($\mu\rightarrow e,\gamma$) is. In the $R$-parametrization $({\bf Y^\dagger_\nu Y_\nu})_{ij}$ has a dependence on $\theta_{13}$, which essentially disappears once the 9-dimensional parameter space is fairly covered.

\item In the $R$-parametrization (which is the most common in the literature) a fair scan implies to allow all the complex $R$-matrices compatible with orthogonality and perturbativity of the Yukawa couplings. The latter requirement has not been properly taken into account in former literature. As a consequence previous scans in the space of the $R$-matrices are typically biassed, since many possible $R$-matrices were excluded from the beginning. 

\item We give a very simple rule to incorporate the perturbativity of Yukawa couplings as a condition in the entries of the orthogonal $R$-matrix. It is given by eq.~(\ref{Pert_2}).
We also give (in Appendix A) an straightforward procedure to completetly scan the space of complex $R$-matrices in a consistent way with this requirement and the orthogonality one.

\item Once such scan is performed the branching ratio BR($\mu\rightarrow e,\gamma$) gets very insensitive to $\theta_{13}$, as already mentioned. Moreover, the values of the branching ratio are typically larger than those quoted in the literature. This comes from the fact that many possibilities that were disregarded from the beginning  (typically with large $R$-entries and thus sizable Yukawas) turn out to be perfectly compatible with the perturbativity constraint, which had not been taken into account.

\item We find this increase of BR($\mu\rightarrow e,\gamma$) a very suggesting result. However, one has to keep in mind that if we scanned the $R-$parameter space in a different way (though still covering the whole space) the relative abundance of points with large BR($\mu\rightarrow e,\gamma$) might change. In Bayesian language, this is equivalent to use a different prior for the parameter space. The dependence of the typical size of BR($\mu\rightarrow e,\gamma$) on the prior is out of the scope of this paper. Nevertheless it is worth noticing that the same prior dependence was implicit in the scatter plots shown in previous literature. 

\item Including leptogenesis constraints (disregarding flavour effects) in the analysis does not introduce any further dependence of BR($\mu\rightarrow e,\gamma$) on $\theta_{13}$. The main impact of leptogenesis, besides remarkably reducing the acceptable volume of the $R$-parameter-space, is a decrease of BR($\mu\rightarrow e,\gamma$) by more than one order of magnitude. This comes from the fact that the efficiency factor, $\eta$, decreases for large $R$-matrix entries. Hence, successful leptogenesis prefers smaller ones, and thus smaller BR($\mu\rightarrow e,\gamma$).

\end{itemize}

\noindent As a concluding remark, scanning the parameter space of the $R-$matrix in the full allowed range is necessary in order to make general statements about predictions of the seesaw scenario. This has not been taken into account in former works, and we consider it an interesting line of work to be explored in forthcoming projects.

\section*{Acknowledgements}

We thank A. Ibarra for very useful discussions. This work has been partially supported by the MICINN, Spain, under contracts FPA-2007--60252 and FPA-2007-60323; Consolider-Ingenio PAU CSD2007-00060 and MULTIDARK CSD2009-00064. We thank as well the Generalitat Valenciana grants  PROMETEO/2009/116 and PROMETEO/2008/069; the Comunidad de Madrid through Proyecto HEPHACOS  ESP-1473 and  the European Commission under contract PITN-GA-2009-237920. B. Zald\'\i var acknowledges the financial support of a FPI (MICINN) grant, with reference BES-2008-004688.

\newpage
\appendix

\section{General scan of the $R-$matrix}
\label{scan}

In this section we explain the details of the scan made on the $R-$matrices, to cover all possibilities compatible with orthogonality,  $R^T R = {\bf 1}$, and the perturbativity condition (\ref{Pert}, \ref{Pert_2}), which for convenience we repeat here:
\begin{equation}
|R_{ij}|^2\lsim \frac{1}{M_i\kappa_j}\ .
\label{PertAp}
\end{equation}
The algorithm presented below has been designed to be easily modified if one considers also  leptogenesis constraints.

In general, given a normal hierarchy among neutrinos, $\kappa_1\ll\kappa_2\ll \kappa_3$, condition (\ref{PertAp}) tells us that the $R-$matrix elements are allowed to be larger (in absolute value) when you move from bottom-right to top-left in the matrix.  Thus,  the element $R_{33}$ ($R_{11}$) presents the smallest (largest)  upper bound. On the other hand, orthogonality implies that, in practice, not all the elements of $R-$matrix can reach their corresponding perturbativity limit. Normally, if one sets any entry of the first row (column), at its maximum magnitude, then the corresponding entries in the same column (row) cannot satisfy orthogonality without violating the perturbativity bound (\ref{PertAp}). Conversely, if the matrix elements $R_{22}, R_{23}, R_{32}, R_{33},$ satisfy their perturbativity constraints, then normally the entries of the first row and column would do it as well. Hence, it makes sense to start imposing perturbativity in the  the bottom-right part of the $R-$matrix. Then the rest is constructed automatically from orthogonality, requiring a final cross-check of the perturbativity condition.

More in detail, using the parametrization  (\ref{R}), we see that
\[
R_{32}^2+R_{33}^2 = c_2^2 ~,
\]
but the perturbativity condition tells us that $R_{32}^2$ is allowed to be much larger than $R_{33}2$, up to phases. So, we can use the perturbativity upper bound on $R_{32}$ to scan $c_2$
\begin{equation}
c_2 = |c_2| e^{i\phi_2}, \hspace{1cm} |c_2|\leq \dfrac{1}{\sqrt{M_3 \kappa_2}}, \hspace{0.5cm} \phi_2\in[0,2\pi].
\label{c2}
\end{equation}
Here the phase $\phi_2$ (and those appearing below) is assumed to be a random number within its interval. Now, for each value of $c_2$ we derive $s_2$ (obviously, up to the sign) and exploit the perturbativity condition on $R_{33}$ to scan $c_1$:
\begin{equation}
c_1 = |c_1| e^{i\phi_1}, \hspace{1cm} |c_1|\leq \dfrac{1}{|c_2|\sqrt{ M_3 \kappa_3}}, \hspace{0.5cm} \phi_1\in[0,2\pi] 
\label{c1}
\end{equation}
(again, for each value of $c_1$ there are two values of $s_1$). At this point we have used the bounds on $R_{32}, R_{33}$, and the orthogonality condition, to scan $\theta_1, \theta_2$.



Now, we have to scan the third complex angle, $\theta_3$. A convenient way to do it is by using the $R_{23}$ element (recall we prefer to impose perturbativity in the bottom-right part of $R$). Thus we scan
\begin{equation}
R_{23} = |R_{23}| e^{i\phi_{23}}, \hspace{1cm} |R_{23}|\leq \dfrac{1}{\sqrt{M_2 \kappa_3}},\hspace{0.5cm} \phi_{23}\in[0,2\pi]   ~.
\label{R23}
\end{equation}
For each value of $R_{23}$ we derive the two possible values of $s_3$:
\begin{equation}
\label{s3}
s_3 = \dfrac{s_2 R_{23} c_1 \pm \sqrt{s_1^2(-R_{23}^2+s_1^2+c_1^2 s_2^2)}}{c_1^2s_2^2+s_1^2} ~.
\end{equation}
Once again, for each value of $s_3$ there are two of $c_3$. Finally, we cross-check for each of these values that the corresponding $R-$matrix is indeed consistent with the perturbativity condition. One can do that directly by using equation (\ref{Pert}) or by checking eq.~(\ref{PertAp}) for the remaining entries.

When leptogenesis constraints are included, most of the initially-allowed values for the first-row entries of $R$, become too big, since they lead to a small efficiency factor, $\eta$ (see sec. \ref{lepto}). In this case, it pays to constrain from the beginning any of the entries in the first row, e.g. $R_{13}$. Then, instead of the previous scan in $R_{23}$, one scans  $R_{13}$ as:
\begin{equation}
R_{13} = |R_{13}| e^{i\phi_{13}}, \hspace{1cm} |R_{13}|\lesssim 1,\hspace{0.5cm} \phi_{13}\in[0,2\pi]   ~.
\label{R13}
\end{equation}
Note that this upper bound on $|R_{13}|$ is normally much lower than its perturbativity bound ($\sim 10^{2}$ in our case), but still is way larger than required by leptogenesis constraints. Now, for each value of $R_{13}$ the value of $s_3$ is given by
\begin{equation}
s_3 = \dfrac{R_{13} s_1 \pm \sqrt{c_1^2 s_2^2(-R_{13}^2+s_1^2+c_1^2 s_2^2)}}{s_1^2+c_1^2 s_22} ~,
\end{equation}
which replaces eq.(\ref{s3}).

As a final comment, recall that the parametrization of $R$ given by eq.~(\ref{R}) is completely general up to reflections changing the sign of det $R$. In our case, however, the scan is completely general since all the relevant physical quantities are invariant under global changes of the sign of  $R$.

\section{Details of the numerical computations}

\subsection{$R-$parametrization}

In this section we explain the strategy for computing the branching ratio BR($\mu\rightarrow e,\gamma$)  for each point in the scan of the  $R-$matrices (see Appendix A).

We have adopted an mSUGRA framework, with universal soft terms at the GUT scale, $M_X$,
\begin{equation}
({\bf m_{L}^2})_{ij} =({\bf m_{e_R}^2})_{ij} = m_0{\bf 1}, \hspace{0.5cm}({\bf A_{e}})_{ij} = A_0 ({\bf Y_e})_{ij},
\end{equation}
where ${\bf m_{L}^2}$, ${\bf m_{e_R}^2}$ and ${\bf A_{e}}$ are the left- and right-handed slepton mass-squared matrices, and the matrix of slepton trilinear couplings. At $M_X$ all the soft terms are diagonal in the basis in which $\bf Y_e$ is diagonal. Below $M_X$, the RG running of ${\bf m_L^2}$ produces off-diagonal entries, due, essentially, to the  contribution proportional to $\bf Y_\nu^\dag Y_\nu$. This constitutes the main source of flavor violation.

We work with our own modified version of the SPheno \cite{Porod:2003um} code. The ${\bf m_L^2}$ matrix is of special interest since, among the slepton mass matrices it is by far the one that develops larger off-diagonal terms. At the seesaw scale, $M$, we evaluate
${\bf m_L^2}$ as
\begin{equation}
{\bf m_L^2} =  {\bf D_{m_L^2}} -\frac{1}{8\pi^2}(3m_0^2+A_0^2) {\bf Y_\nu^\dag D_L Y_\nu }  ~.
\label{mL2}
\end{equation}
Here ${\bf D_{m_L^2}}$ is the result of running ${\bf m_L^2}$ from $M_X$ until $M$, switching off the contribution from neutrino Yukawa couplings. The second term in (\ref{mL2}) is the contribution coming from the neutrino Yukawas, evaluated at the leading-log approximation. This contribution contains the off-diagonal entries of ${\bf m_L^2}$. The value of ${\bf Y_\nu}$ at the $M-$scale is obtained from the $R-$parametrization formula (\ref{param1}) for each point in the scan of the  $R-$matrices. Finally, ${\bf m_L^2}$ is run down to low-energy (neutrino Yukawas do not play any role in this RG-interval since right-handed neutrinos are decoupled).

The rest of physical quantities (charged slepton mass matrices, gauge couplings, GUT scale, charged lepton yukawas, etc.) are taken directly from SPheno, which imposes the $M_X$ scale to be the one where  gauge couplings unify. We also extracted from SPheno the parameters of the neutralinos and charginos. Finally, we followed ref. \cite{Hisano:1995cp} (implemented in SPheno) to calculate the branching ratio at 1-loop level.

\subsection{$V_L-$parametrization}

For the $V_L-$parametrization, we have used the SPheno code as well. However, the original code is not prepared to introduce the initial parameters according to this parametrization. In particular, the original version works with given values of right-handed neutrinos masses, $M_i$,  from the beginning. But in the  $V_L-$parametrization, $M_i$ (and $V_R$) are obtained at the seesaw scale, $M$ (a suitable average of $M_i$), from eq.~(\ref{VRM}). On the other hand, the neutrino Yukawa eigenvalues, $D_Y$ and the $V_L$ matrix are indeed initial parameters, defined at the high-scale (and these are the ones in which we perform our scan of the parameter space). 

Consequently, we modified the code, incorporating an iterative procedure to determine $M_i$ in a consistent way with all the boundary conditions (at low- and high-scale). In this way, the full ${\bf m_L^2}$ matrix is obtained directly from SPheno, and can be used to the computation of BR$(\mu\rightarrow e,\gamma)$.

\pagebreak

\bibliographystyle{JHEP}    
\bibliography{cmrrz_refs}	 

\end{document}